\documentclass[aps,prb,reprint,amsmath,amssymb,groupedaddress,notitlepage]{revtex4-2}
\usepackage[utf8]{inputenc}
\usepackage{bm}
\usepackage{enumerate}
\usepackage{graphicx}
\usepackage{cancel}
\usepackage{xcolor}
\usepackage{verbatim}
\usepackage{hyperref}
\hypersetup{colorlinks=true,
linkcolor=blue,
citecolor=blue,
urlcolor=blue,
filecolor=blue
}

\def\bra#1{\langle{#1}|}
\def\ket#1{|{#1}\rangle}
\def\braket#1{\langle{#1}\rangle}
\newcommand{\bmk}[3]{\langle #1 | #2 | #3 \rangle}

\newcommand{\uu}{\left|\uparrow\uparrow\right\rangle}
\newcommand{\ud}{\left|\uparrow\downarrow\right\rangle}
\newcommand{\du}{\left|\downarrow\uparrow\right\rangle}
\newcommand{\dd}{\left|\downarrow\downarrow\right\rangle}
\newcommand{\sing}{\left|S_{02}\right\rangle}

\begin{abstract}
Electric dipole spin resonance (EDSR) is a commonly used tool for manipulation and spectroscopy of quantum-dot-based spin qubits.
When an EDSR experiment is embedded in a transport setup and Pauli spin blockade is used as means for spin-state read-out, then measured resonant responses in the leakage current indeed carry information about the level structure of the system under study.
However, the actual line shape of these current resonances differs substantially from experiment to experiment, varying from being symmetric to asymmetric and from being a peak to a dip, a thorough understanding of which is still lacking.
Here, we investigate theoretically the detailed line shape of EDSR-induced resonances in the leakage current in the regime of spin blockade, and we connect different line shapes to the different underlying physical mechanisms that can enable the EDSR.
We carry out both numerical and analytical investigations, producing simple analytic expressions that give insight in the physics at play.
Our results thus provide a means to extract more information about the detailed system parameters of quantum dots hosting spin qubits from an EDSR experiment than just their level structure based on the location of the resonances.
\end{abstract}

\begin{document}

\title{Line Shapes of Electric Dipole Spin Resonance in Pauli Spin Blockade}

\author{Arnau Sala}
\altaffiliation{Current address: Imec, Kapeldreef 75, 3001 Heverlee, Belgium. \\
Department of Electrical Engineering, KU Leuven, Kasteelpark Arenberg 10, 3001 Leuven, Belgium.}
\affiliation{Center for Quantum Spintronics, Department of Physics, Norwegian University of Science and Technology, NO-7491 Trondheim, Norway}

\author{Jeroen Danon}
\affiliation{Center for Quantum Spintronics, Department of Physics, Norwegian University of Science and Technology, NO-7491 Trondheim, Norway}

\date{\today}

\maketitle

\section{Introduction}

Semiconductor spin qubits and other spin-based nanostructures require accurate mechanisms for spin manipulation~\cite{DiVincenzo2000,Ladd2010}.
Spin qubits are encoded in the spin states of electrons or holes localized in quantum dots, and manipulation of their quantum state can thus most straightforwardly be achieved by implementing some form of local spin resonance~\cite{Wiel2003,Hanson2007,Zwanenburg2013}. 
Since applying strongly localized oscillating magnetic fields is very challenging in practice~\cite{Koppens2006}, a broadly used method for manipulation is electric dipole spin resonance (EDSR)~\cite{Golovach2006,Bulaev2007}, where via intrinsic or artificial~\cite{Tokura2006,Zhang2021} spin-orbit coupling the spins can be controlled using oscillating \textit{electric} fields.
Indeed, EDSR is a well-established tool to manipulate electron~\cite{Nowack2007,Pioro-Ladriere2008,Nadj-Perge2010,Pei2012,Yoneda2014,Noiri2016,Corna2018}
and hole~\cite{Maurand2016,Crippa2018,Pribiag2013,Watzinger2018,Hendrickx2020,Froning2021}
spin states in quantum dots.

A closely related and widely used application of EDSR in quantum dots is to use it as a spectroscopic tool, to access intrinsic parameters of the system.
Indeed, apart from yielding coherent spin rotations useful for quantum-information applications, the EDSR response of a system can be mapped out as a function of, e.g., driving frequency and applied magnetic field, which allows to infer system parameters such as the effective $g$-factors on the quantum dots or characterize the effects of hyperfine interaction and spin-orbit interaction in the system~\cite{Pfund2007,Pfund2007a,Schroer2011,Nadj-Perge2012,Frolov2012}.

EDSR-induced spin rotations, whether used for coherent manipulation or spectroscopy, have to be detected in the end by some type of spin-to-charge conversion.
A commonly used tool for this is Pauli spin blockade in a multi-dot setup:
The EDSR experiment is concluded with inducing an interdot charge tunneling event that is spin selective by virtue of the Pauli principle~\cite{Ono2002}.
Embedding this tunneling in a transport setup, an EDSR response can then be detected in the form of a resonant response in the current~\cite{Nadj-Perge2012}.
In the context of spectroscopy, the location of this current response as a function of driving frequency thus provides information about the level structure of the system.

However, EDSR spectroscopy experiments often show many intricate details in the data: variations in width and line shape of the resonant response across parameter space, as well as changes in sign of the response, i.e., both resonant peaks and dips have been observed in the same experiment~\cite{Laird2009,Stehlik2014}.
Although such features must contain additional information about the physics underlying the EDSR, such as the detailed nature of the spin-orbit interaction in the system, the exact manifestation of the resonant EDSR response in Pauli spin blockade still lacks a thorough theoretical understanding.

Here we investigate the EDSR line shape in Pauli spin blockade in detail, with the aim to shed more light on the physics underlying EDSR in quantum dots.
We identify different ways in which various types of (effective) spin-orbit coupling can leverage electric driving into an effective oscillating magnetic field acting on the spins of the localized carriers.
We establish a connection between each of these mechanisms and qualitative features of the EDSR line shape, which thus can be used to disentangle the contributions of different types of spin-orbit interaction in experiment, thereby providing more detailed understanding of the spin physics in the system.

The rest of this paper is organized as follows.
In Sec.~\ref{sec:edsr} we briefly explain the basics of EDSR in Pauli spin blockade and identify the main mechanisms responsible for it.
In Sec.~\ref{sec:model} we introduce the model we use to describe the EDSR, show how the different mechanisms we consider can be incorporated, and proceed by reducing it to the simplest minimal model that captures all essential physics.
Sec.~\ref{sec:numerics} presents numerical simulations that show how the line shape of the resonant response can indeed differ qualitatively, ranging from symmetric to asymmetric and from peak to dip, depending on the mechanism at play and the details of the interdot coupling.
We then proceed in Sec.~\ref{sec:anal} to derive approximate analytic expressions for the resonant response of the leakage current in the weak-driving limit, covering all mechanisms considered, and show how the analytic results match the numerical simulations.
Understanding the structure of our analytic results finally allows us to connect specific features in the EDSR line shape to the underlying mechanisms at play, as we discuss in Sec.~\ref{sec:conclusion}.

\section{EDSR in Pauli spin blockade}\label{sec:edsr}

\begin{figure}
  \centering
  \includegraphics[width=\linewidth]{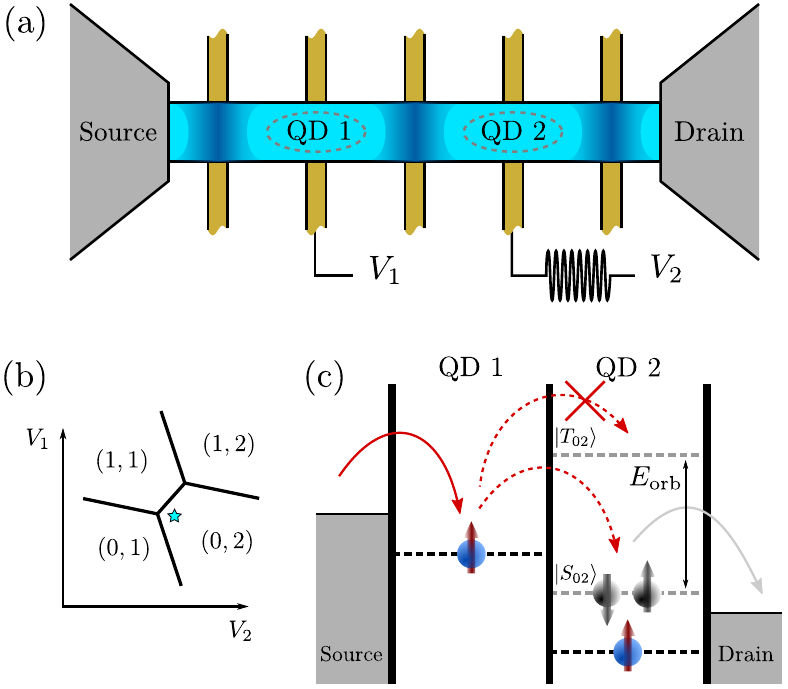}
    \caption{(a) A double-dot structure can be created inside a semiconductor nanowire by depositing the wire on top of a set of electric gates that can then be used to create tunnel barriers and to control the electrostatic potentials on the dots.
    	(b) Part of a typical double-dot charge stability diagram.
    	The system is tuned close to the (1,1)--(0,2) transition, e.g., at the point marked by the blue star.
    	(c) Simplified level structure of the states involved in transport, at zero magnetic field.
    	In the spin-blockade regime, the only accessible (0,2) state is a spin singlet; the other (0,2) states are separated by a large orbital energy $E_\text{orb}$.\label{fig:dots}}
\end{figure}

We study a system composed of two quantum dots in a host material where the relevant carriers, electrons or holes, experience strong spin-orbit interaction (SOI), such as InAs, InSb, Si, or Ge~\cite{Pfund2007,Fasth2007,Ferdous2018,Jock2018,Corna2018,Harvey-Collard2019,Watzinger2018,Hendrickx2020,Froning2021,Froning2021a,Hao2010,Pribiag2013}.
The dots are tunnel coupled to each other and to source and drain reservoirs, see Fig.~\ref{fig:dots}(a) where we assumed a nanowire-based setup.
The system is then tuned to a regime of Pauli spin blockade, for instance close to the (1,1)--(0,2) ground state charge transition [where $(n,m)$ denotes the state with $n(m)$ excess carriers in the left(right) dot], see Fig.~\ref{fig:dots}(b).
In the presence of a large enough bias voltage, the transport cycle $(0,1) \to (1,1) \to (0,2) \to (0,1)$ is then in principle allowed.
However, due to the Pauli principle and the large on-site orbital energy splitting, the transition $(1,1) \to (0,2)$ is spin-selective, such that the transport cycle can only be completed if the (1,1) state is a spin singlet.
Tunneling from (0,1) into one of the (1,1) triplets will lead to blockade of the current, as illustrated in Fig.~\ref{fig:dots}(c).

In order to escape from the blockade and contribute to a finite leakage current a spin flip is required.
At small magnetic field, such spin mixing can be provided by the hyperfine coupling to the fluctuating nuclear spin bath in materials with a significant fraction of spinful nuclei~\cite{Koppens2005,Jouravlev2006}; increasing the magnetic field to larger values will enable the SOI to contribute to the spin mixing, by making the tunneling effectively non-spin-conserving~\cite{Danon2009a}.
Indeed, in materials with strong SOI the leakage current is usually significant, showing a dip-like structure around zero magnetic field~\cite{Pfund2007a,Nadj-Perge2010a,Li2015,Bohuslavskyi2016,Zarassi2017}.

A typical EDSR-based spectroscopy experiment is performed by additionally applying an oscillating voltage to one of the gates controlling the on-site dot potentials, such as shown in Fig.~\ref{fig:dots}(a).
Measuring the leakage current as a function of applied magnetic field and frequency of the driving voltage usually yields resonant responses in the current~\cite{Laird2009,Nadj-Perge2010,Schroer2011,Nadj-Perge2012,Frolov2012,Pribiag2013}, which allow to extract the field-dependent level structure of the (1,1)--(0,2) subspace.
However, the detailed manifestation of the current resonances as a function of the driving frequency, driving power, and magnetic field differs a lot from one experiment to another, 
ranging from dips to peaks, with varying line shapes~\cite{Stehlik2014,Crippa2018}.

In order to develop a detailed understanding of the resonant EDSR response in the leakage current, we consider several possible underlying mechanisms:
(\textit{i}) Assuming a driving on one of the gates that controls the on-site electrostatic potential in one of the dots, such as sketched in Fig.~\ref{fig:dots}(a), the oscillating signal yields a periodic modulation of the energy splitting between the (1,1) and (0,2) charge states~\cite{Danon2014,Stehlik2016}.
Exchange effects that are mediated by the SOI-induced non-spin-conserving tunnel coupling can effectively translate this modulation into oscillating spin-dependent couplings within the (1,1) subspace, which, when on resonance, can induce spin flips.
(\textit{ii}) The periodic change in the potential landscape due to the oscillating gate voltage also results in the wave functions of the localized carriers to change periodically.
Combined with the strong SOI, this translates directly into a periodic effective magnetic field coupling to the spin of the carriers, enabling a spin resonance~\cite{Golovach2006,Nowack2007}.
The magnitude and direction of this field depend on the details of the SOI.
(\textit{iii}) Another similar possible mechanism is based on the hyperfine coupling of the carriers to the randomly polarized nuclear spin ensembles in the two dots, relevant for host materials with significant fractions of spinful nuclei.
To good approximation, this coupling results in an effective (random) quasistatic ``nuclear field'' coupling to the spin of the carriers, its details depending on the actual spin polarization of the nuclei and the carrier wave function.
A periodic change of that wave function will thus lead to a periodically changing nuclear field, which can also induce spin rotations~\cite{Rudner2007}.
(\textit{iv}) Finally, also for the case of artificial SOI induced by a slanting magnetic field~\cite{Tokura2006} a small periodic change of the carriers' location directly translates to an oscillating magnetic field acting on their spin.

Altogether, we thus need to consider the effects on the spin blockade of (1) a periodic modulation of the on-site potentials combined with SOI and (2) an oscillating effective on-site magnetic field.
Previous similar theoretical studies of EDSR in the spin-blockade regime focused on mechanism (\textit{i}) in the strong-driving limit~\cite{Danon2014,Stehlik2016,Giavaras2019} or on mechanism (\textit{ii}) in the absence of both SOI and coherent interdot coupling~\cite{Bobbert2007,Roundy2013,Malla2017,Jamali2021}.
Other related studies focus on a lifting of the Pauli spin blockade in the presence of SOI and hyperfine coupling but without an external driving of the system~\cite{Mutter2020,Szechenyi2017}.
Here, we want to focus on the weak-driving limit, where the amplitude of the driving-induced effective oscillating magnetic field in the (1,1) subspace is always smaller than the level splittings we want to probe.

\section{Model}\label{sec:model}

\textit{Hamiltonian}---Since all relevant dynamics take place in the (1,1)--(0,2) subspace, we only need to consider the four (1,1) spin states, for which we use the basis $\{ \uu, \ud, \du, \dd\}$, and the (0,2) singlet state $\sing$.
Choosing the spin quantization axis along the direction of the magnetic field we have $\mathbf B = B \hat z$, and assuming that the quantum dots have different $g$-factors $g_i$ due to the strong SOI, we thus write the Hamiltonian as:
\begin{align}\label{eq:ham5}
	H_5  = {}&{} \left(
	\begin{array}{c c c c c}
		B_s    &     0    &     0   &     0    &     t^*_1 \\
		0     &    B_a    &     0    &   0    &     t_2 \\
		0    &     0    &     -B_a    &    0    &     -t^*_2 \\
		0    &   0    &    0    &    -B_s    &     t_1 \\
		t_1    &     t^*_2    &     -t_2    &     t^*_1    &     -E_0
	\end{array}
	\right),
\end{align}
where we have defined $B_s \equiv \tfrac{1}{2}(g_1 + g_2) \mu_{\rm B} B$ and $B_a \equiv \tfrac{1}{2}( g_1- g_2) \mu_{\rm B} B$, with $\mu_{\rm B}$ the Bohr magneton, and we did not include the effects of the driving yet.
All relevant details of the electrostatic tuning of the double dot are captured in $E_0$, which describes the energy splitting between the (1,1) and (0,2) charge states in a Hubbard-like description of the system~\cite{Burkard1999,DasSarma2011,Medford2013a,Taylor2013}.
One effect of SOI is also included in this Hamiltonian:
The tunneling elements $t_i$ can account for both spin-conserving and spin-flip interdot tunneling.
This allows for transitions from the charge state (1,1) to (0,2) even when the carriers in the (1,1) state do not form a spin singlet.
In the absence of SOI, we would write $t_1=0$ and assume $t_2$ to be real.

The source and drain leads are electron reservoirs, the effect of which can be included in the five-level dynamics described by $H_5$ via an escape rate $\Gamma$:
When the system is in the state $\sing$, one carrier leaves the doubly occupied dot to the drain at a rate $\Gamma$ and subsequently one of the four (1,1) states is refilled from the source, with equal probabilities.
In this picture, the leakage current through the system is given by $I = \Gamma P_{S_{02}}$, where the carrier charge has been set to 1 and $P_{S_{02}}$ is the average probability for the system to be in the state $\sing$.

\begin{figure}[t]
	\centering
	\includegraphics[width=\linewidth]{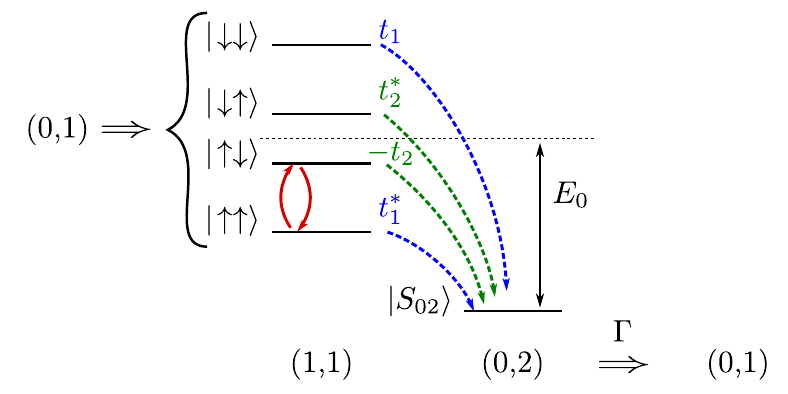}
	\caption{Level structure of the relevant (1,1) and (0,2) states in the presence of a magnetic field.
		Due to spin-orbit interaction, all four (1,1) eigenstates are coupled to the (0,2) singlet.
		The system is driven with an oscillating voltage applied to one of the gates, see Fig.~\ref{fig:dots}(a), which can be tuned into resonance with one of the single-spin transitions, such as the one on the right dot, indicated in green. \label{fig:levels}}
\end{figure}

In Fig.~\ref{fig:levels} we show a sketch of the level structure and the tunneling processes we thus describe.
In the presence of a magnetic field, the four (1,1) basis states are split in energy and all four are coupled to $\sing$, which decays to (0,1) with a rate $\Gamma$, after which one of the (1,1) states is immediately refilled again.
The resonances that are usually probed in experiment are within the (1,1) subspace, such as the one indicated in red which corresponds to a spin resonance on the second dot.

\textit{Driving}---We include the electric driving of the system by adding a time-dependent term to the Hamiltonian,
\begin{align}\label{eq:h5a}
  H_5^{(\alpha)}(t) = {}&{} H_5 + A \cos(\omega t) V_\text{drive}^{(\alpha)},
\end{align}
where $A$ and $\omega$ characterize the amplitude and frequency of the driving, respectively.
The subscript $\alpha \in \{ \text{d,z,x} \}$ labels the different driving mechanisms we investigate:
\begin{align}
  V_\text{drive}^{(\text{d})} = {}&{} |S_{02}\rangle\langle S_{02} |, \label{eq:vd}\\
  V_\text{drive}^{(\text{z})} = {}&{} 
      \left| \uparrow\uparrow\rangle\langle \uparrow\uparrow \right|
    - \left| \uparrow\downarrow\rangle\langle \uparrow\downarrow \right|
    + \left| \downarrow\uparrow\rangle\langle \downarrow\uparrow \right|
    - \left| \downarrow\downarrow\rangle\langle \downarrow\downarrow \right|, \label{eq:vz}\\
  V_\text{drive}^{(\text{x})} = {}&{} 
      \left| \uparrow\uparrow\rangle\langle \uparrow\downarrow \right|
    + \left| \uparrow\downarrow\rangle\langle \uparrow\uparrow \right|
    + \left| \downarrow\uparrow\rangle\langle \downarrow\downarrow \right|
    + \left| \downarrow\downarrow\rangle\langle \downarrow\uparrow \right|.\label{eq:vx}
\end{align}
The first term, Eq.~(\ref{eq:vd}), results in $E_0 \to E_0 - A\,\cos(\omega t)$ and thus incorporates mechanism (\textit{i}) into our Hamiltonian, where the EDSR is driven by an oscillating interdot detuning.
The other mechanisms rely on an oscillating effective magnetic field on the dots, in some cases randomly oriented, which we can include via the second or third term, Eqs.~(\ref{eq:vz},\ref{eq:vx}).
In both these terms we focused on the oscillating field that couples to the second dot, but the same theory can be used to describe a resonance on the first dot, simply by swapping the dot labels
The two terms allow us to investigate the difference between the oscillating field being parallel or perpendicular to the applied static field, where we consider a perpendicular field along the $x$-direction, without loss of generality.

\textit{Reduction to a three-level problem}---Due to the strong SOI, the two dots have typically substantially different effective $g$-factors~\cite{Nadj-Perge2010,Schroer2011,Nadj-Perge2012,Stehlik2014}.
We thus assume the energy levels of the four (1,1) basis states used in Eq.~(\ref{eq:ham5}) to be well separated and, therefore, we can study each of the single-spin resonances independently.
This allows us to reduce the dimensionality of the Hamiltonian from five to three, while still capturing all the essential physics of the specific resonance under investigation~\cite{Danon2014,Stehlik2016}.
Depending on which spin resonance we want to describe, the reduced three-dimensional basis reads $\{ \uu, \ud, \sing \}$, $\{ \ud, \dd, \sing \}$, $\{ \uu, \du, \sing \}$, or $\{ \du, \dd, \sing \}$, and in all cases the projected three-level Hamiltonian becomes
\begin{align}\label{eq:ham3}
	H_3 = {}&{}
	\left( \begin{array}{c c c}
		B   &     0     &     q_1   \\
		0   &     -B    &     q_2   \\
		q_1^* &     q_2^*   &     -E_0
	\end{array} \right),
\end{align}
where $B$, $q_1$ and $q_2$ are different for each of the four possible subspaces.
Labeling the basis states used in (\ref{eq:ham3}) as $\{\ket{1}, \ket{2}, \ket{S} \}$, we see that the three driving terms of interest now read $V_\text{drive}^{(\text{d})} = \ket{S}\bra{S}$, $V_\text{drive}^{(\text{z})} = \ket{1}\bra{1} - \ket{2}\bra{2}$, and $V_\text{drive}^{(\text{x})} = \ket{1}\bra{2} + \ket{2}\bra{1}$, and we can construct our full time-dependent three-level Hamiltonian as
\begin{align}\label{eq:h3a}
	H_3^{(\alpha)}(t) = {}&{} H_3 + A \cos(\omega t) V_\text{drive}^{(\alpha)}.
\end{align}
We embed this subsystem in a transport setup in a similar way as before:
We assume that the state $\ket{S}$ decays with a rate $\Gamma$, after which one of the two (1,1) states is refilled with equal probability.

The general approach to find the average current through the driven system will be to find the steady-state solution of a Lindblad equation for the density matrix,
\begin{align}\label{eq:lind3}
  \frac{d\rho}{dt} = {}&{} -i[H_3^{(\alpha)}(t),\rho] + \Gamma[\rho],
\end{align}
where we have set $\hbar = 1$ and
\begin{align}
 \Gamma[\rho] = {}&{} \frac{1}{2} \Gamma
 \left( \begin{array}{c c c}
  \rho_{ss}       &     0     &     - \rho_{1s}   \\
  0               & \rho_{ss} &     - \rho_{2s}   \\
  -\rho_{s1}      & -\rho_{s2} &     - 2 \rho_{ss}
  \end{array} \right),
\end{align}
which describes the decay of the state $\ket{S}$ to the drain at a rate $\Gamma$ with the consequent annihilation of the coherences $\rho_{is}$, and the subsequent refilling of the two (1,1) states.
The steady-state solution for $\rho_{ss}$ gives the average occupancy of the (0,2) charge state and, thus, the leakage current via $I = \Gamma \rho_{ss}$.

Consistent with most experiments, we will assume the energy $\Gamma$ to be larger than all other energy scales except $E_0$, which is allowed to be comparable to $\Gamma$.
Based on our assumption of weak driving, which amounts to the regime $A \ll \omega$ (for ``magnetic'' driving) or $A \ll \omega \Gamma^2/q^2$ (for ``detuning'' driving), we will restrict our investigations to the first harmonic, i.e., where the driving frequency matches the energy difference between the two (1,1) states, $\omega \approx 2 B$.

We will now first present numerical evaluations of the average leakage current, based on solving Eq.~(\ref{eq:lind3}) in steady-state, for the three different driving terms.
Then we will proceed to derive approximate analytic expressions for the leakage current for the three types of driving, and compare those to the numerical results.

\section{Numerical results}\label{sec:numerics}

Due to the oscillating nature of the Hamiltonian, a true equilibrium solution to the Lindblad equation does not exist.
Instead, we look for a time-averaged steady-state solution.
To this end we expand the density matrix in Fourier modes $\rho (t) = \sum_n \rho^{(n)} e^{in\omega t}$, resulting in
\begin{align}\label{eq:fourier}
\sum_{n} e^{i n \omega t} \frac{d}{d t} \rho^{(n)}  ={}&{} \sum_{n}  e^{i n \omega t}\left\{ -i[H_3^{(\alpha)}(t),\rho^{(n)}]  \right. \notag \\
      {}&{} \quad\quad\quad\left. + \Gamma[\rho^{(n)}] - in\omega\rho^{(n)} \right\}.
\end{align}
The harmonic time dependence of $H_3^{(\alpha)}$ transforms into a time-independent coupling between $\rho^{(n)}$ and $\rho^{(n\pm 1)}$, thereby in principle facilitating an evaluation of the Fourier components $\rho^{(n)}$ in equilibrium by solving $d\rho^{(n)}/dt=0$.
The time-averaged steady-state density matrix is then $\rho^{(0)}$ and the current follows as $I=\Gamma \rho^{(0)}_{ss}$.

For our numerical simulations, we truncate Eq.~(\ref{eq:fourier}) at $n = \pm 10$,
after making sure that the results do not change significantly when more modes are included.
The resulting set of equations is supplemented with the boundary conditions ${\rm Tr}\{ \rho^{(0)}\} = 1$ and ${\rm Tr}\{ \rho^{(n)}\} = 0$ for all $n\neq 0$ and can then easily be solved.
We investigate the resulting current profile for the three different EDSR mechanisms as a function of the driving frequency $\omega\approx 2B$ for fixed values of the external magnetic field and the amplitude of the driving.
We note here that, although we focus in this work on the single-photon resonance, our simulations also reproduce the behavior of the multi-photon resonances observed in Ref.~\cite{Stehlik2014} and explained in~\cite{Danon2014}, including the alternating sign of the response.

\begin{figure}
  \centering
  \includegraphics[width=\linewidth]{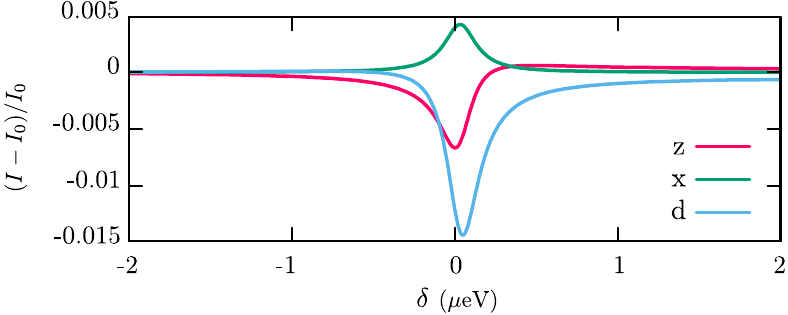}
  \caption{Numerically calculated current resonances for the three different  EDSR mechanisms.
  	We plot $\delta I = (I - I_0)/I_0$ as a function of the frequency detuning $\delta = \omega - 2B$, i.e., we subtract the background current $I_0$ far away from the resonance and normalize the signal to $I_0$.
  	We further used $E_0 = 50\,\mu$eV, $q_1 = 3\,\mu$eV, $q_2 = 2\,\mu$eV, $B = 3\,\mu$eV and  $\Gamma = 150\,\mu$eV.
  	We set the driving amplitude $A = 0.5\,\mu$eV for z-driving, $A = 0.02\,\mu$eV for x-driving, and $A = 20\,\mu$eV for d-driving.\label{fig:num}}
\end{figure}

In Fig.~\ref{fig:num} we show typical numerical results for the three mechanisms.
In the Figure we indeed see both resonant peaks and dips appearing in the current, similar to what is typically observed in experiment.
We further note that different driving mechanisms give rise to qualitatively different resonant line shapes:
We find an asymmetric dip in the current when the driving effectively couples to the (1,1) states via $B_z$, a similar but reversed asymmetric dip when the driving couples mostly to $E_0$, and a peak in the current when the driving couples via $B_x$.
These numerical results will serve to assess the validity of the analytical expressions obtained below.
The following sections contain a further analytic investigation of the shape of the current profile and its dependence on the type of driving and the system parameters.

\section{Analytical results}\label{sec:anal}

The method with which we choose to evaluate the current analytically is different for the cases of detuning driving and magnetic driving.
The reason is that the oscillations in the effective SOI- or hyperfine-induced magnetic fields due to the periodic electric detuning of the dots are typically much smaller than $B$.
This allows us to start with separating time scales, making use of the assumption that $\Gamma$ is large, and isolate the slow dynamics in the (1,1) subspace, to which we can apply perturbation theory in $A/\omega$.
In contrast, when driving the detuning, the oscillating terms that realize spin rotations in the (1,1) space arise from exchange coupling to $\ket{S}$, which relies on coherence between the two subspaces; separating time scales is thus not an option in this case.

\subsection{Effective magnetic driving}

We thus start in this case by separating the time scales of the ``slow'' $2\times 2$ block of the density matrix describing the (1,1) subspace and the other ``fast'' elements, which evolve on the time scale $\Gamma^{-1}$.
Assuming that the five elements involving $\ket{S}$ reach steady state instantaneously on the slow time scale, we can integrate out their dynamics, resulting in a Lindblad equation for the $2\times 2$ subspace.
Introducing a matrix notation, we write
\begin{align}\label{eq:rho2}
  \frac{d\boldsymbol{\rho}}{dt} = {}&{} \left[ \mathcal{M} + A \cos (\omega t)\, \mathcal{V}_\text{drive}^{(\alpha)} \right] \boldsymbol{\rho},
\end{align}
where the density matrix is now written as a vector,
\begin{align}
  \boldsymbol{\rho} = {}&{} \left( 
\rho_{11} ~ \rho_{12} ~ \rho_{21} ~ \rho_{22}
\right)^T.
\end{align}
The $4\times 4$ matrix $\mathcal{M}$ contains the time-independent part of the Lindblad equation and $A \cos(\omega t)\,\mathcal{V}_\text{drive}^{(\alpha)}$, with $\alpha \in \{ {\rm z},{\rm x}\}$, contains the terms due to the driving.
We refer to App.~\ref{app:magdrive} for an explicit expression for the matrices used.

We then expand the density matrix again in Fourier modes, $\boldsymbol \rho(t) = \sum_n e^{in\omega t}\boldsymbol \rho^{(n)}$, which, after dividing out common factors of $e^{in\omega t}$, yields
\begin{align}\label{eq:rho2F}
	\frac{d\boldsymbol{\rho}^{(n)}}{dt} = {}&{} [ \mathcal{M} - in\omega]\boldsymbol{\rho}^{(n)}
	 + \frac{A}{2}\,\mathcal{V}_\text{drive}^{(\alpha)}[ \boldsymbol{\rho}^{(n-1)} + \boldsymbol{\rho}^{(n+1)}].
\end{align}
After truncating this expression at some maximal $\pm n$, it can be solved to find the steady-state solution for the Fourier components of $\boldsymbol \rho$.
From this solution the corresponding fast element $\rho^{(0)}_{ss}$ can be calculated, and the leakage current follows again as $I=\Gamma \rho_{ss}^{(0)}$
For our analytic investigations we will make use of the weak-driving assumption $A \ll \omega$, which suggests that including only the lowest Fourier components $n = -1, 0, 1$ can already yield a good result.

\subsubsection{Oscillating effective field along $\hat x$}\label{sec:bx}

We first explore the behavior of the current when the driving results in an oscillating effective magnetic field along a perpendicular direction, and without loss of generality we take the $x$-direction.
Using ${\cal V}^{(x)}_\text{drive}$ in Eq.~(\ref{eq:rho2F}) we then find for the leakage current close to resonance the following approximate expression, valid in the limits $\Gamma \gg A, B, q_{1,2}$ and $\omega\gg A$:
\begin{align}\label{eq:Ibx}
	\frac{I}{I_0} = 1 +  \left( \frac{q_1^2 - q_2^2}{2 q_1 q_2} \right)^2 \frac{A^2}{ A^2 + \gamma^2 + (\delta + \epsilon_2 - \epsilon_1)^2},
\end{align}
where $\delta = \omega -2B$ and we introduced
\begin{align}
	\gamma = \frac{2\Gamma (q_1^2 + q_2^2)}{4 E_0^2 + \Gamma^2},
\end{align}
the average effective escape rate from the two (1,1) states.
The background leakage current is approximately
\begin{align}
I_0 = \gamma \left(\frac{2 q_1q_2}{q_1^2 + q_2^2} \right)^2,
\end{align}
and results from only considering the effective escape rates from $\ket 1$ and $\ket 2$ via $\ket S$, which are approximately $\gamma_{1,2} = 4q^2_{1,2}\Gamma / (4E_0^2 + \Gamma^2)$.
Note how for symmetric couplings $q_1 = q_2$ the background current is simply $\gamma$.

From Eq.~(\ref{eq:Ibx}) we see that in this case the response always appears as a peak in the current.
Indeed, the driving term in the Hamiltonian acts along $\sigma_x$ in the two-level (1,1) subspace, resulting in a resonant Rabi driving of the two states.
If the escape rates $\gamma_1$ and $\gamma_2$ differ significantly, then a driving-induced mixing of the two states will always lead to a slightly enhanced current by reducing the importance of the ``bottleneck'' state with the lowest escape rate.
For this reason the resonant response scales with $(q_1^2-q_2^2)^2 \propto (\gamma_1-\gamma_2)^2$.
The width of the peak contains a contribution from the power broadening due to the Rabi driving as well as from the life-time broadening of the two levels.

\begin{figure}[t!]
  \centering
  \includegraphics[width=\linewidth]{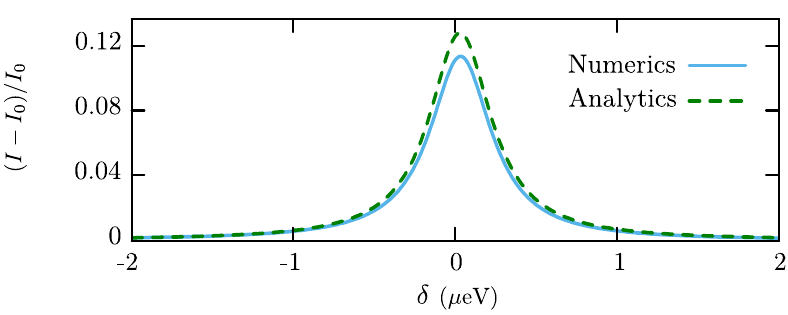}
  \caption{Current profile for a driving in the magnetic field along the $x$ direction. The dashed green curve is the solution from Eq.~(\ref{eq:Ibx}). The solid blue curve is the numerical solution. In this figure we have used a larger amplitude than in Fig.~\ref{fig:num}, $A = 0.2\mu$eV, to better display the shape of the curve. The rest of the parameters are: $E_0 = 50\,\mu$eV, $q_1 = 3\,\mu$eV, $q_2 = 2\,\mu$eV, $B = 3\,\mu$eV and  $\Gamma = 150\,\mu$eV. \label{fig:bx}}
\end{figure}
In Fig.~\ref{fig:bx} we compare our approximate analytic expression to the numerical results obtained before and we see that the simple expression (\ref{eq:Ibx}) captures both the peak height and line width reasonably well.


\subsubsection{Oscillating effective field along $\hat z$}

Now we switch to driving that effectively results in an oscillating field along $\hat z$.
Using the same approximations and notation as before, we then find for the leakage current close to resonance
\begin{align}\label{eq:Ibz}
	\frac{I}{I_0} = {}&{}
	1 - \frac{A^2}{4B^2} \frac{\gamma^2 +(\epsilon_2-\epsilon_1)^2 + 2(\epsilon_2-\epsilon_1) \delta }{\gamma^2+ (\delta + \epsilon_2 - \epsilon_1)^2}, 
\end{align}
where we further introduced the energies
\begin{align}
	\epsilon_{1,2} = \frac{4 E_0 q_{1,2}^2}{4 E_0^2 + \Gamma^2},
\end{align}
which are the exchange-induced shifts of the (1,1) levels.

\begin{figure}
	\centering
	\includegraphics[width=\linewidth]{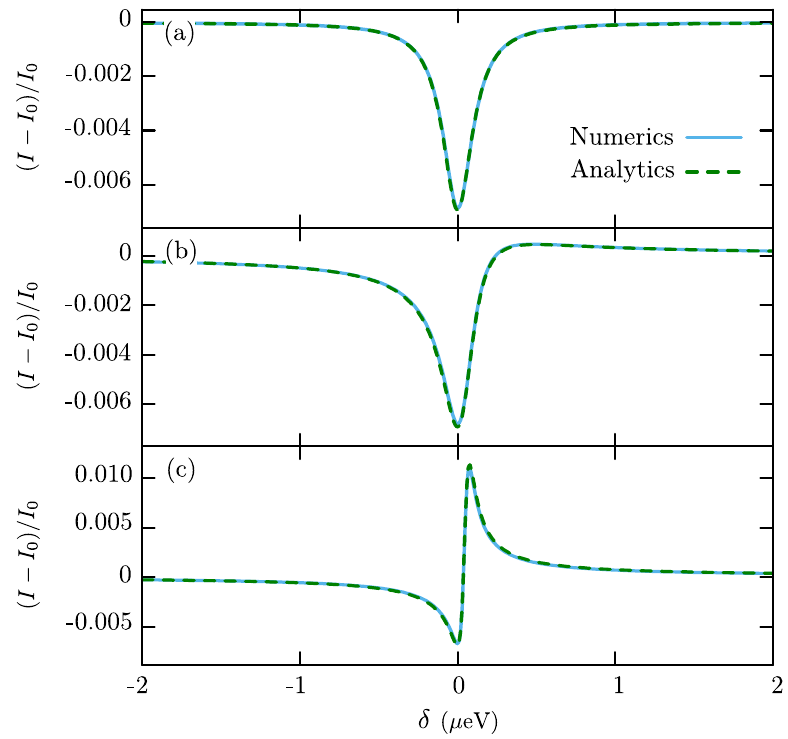}
	\caption{Current profile for a driving in the magnetic field along the $z$ direction with an electrostatic offset $E_0 = 50\,\mu$eV. The green dashed curves are obtained from Eq.~(\ref{eq:Ibz}). Numerical results (solid blue) are added for comparison. (a) For equal couplings $q_1 = q_2 = 2.5\,\mu$eV we observe a symmetric dip in the current. (b) When the couplings are different, $q_1 = 3\,\mu$eV and $q_2 = 2\,\mu$eV, the dip shows a slight asymmetry. (c) Increasing the energy $E_0 = 120\,\mu$eV  and further biasing the couplings, $q_1 = 3\,\mu$eV and $q_2 = 1\,\mu$eV, creates an asymmetric shape in the current with a peak that is more prominent than the dip. For this plot we have used the same parameters as in Fig.~\ref{fig:num}: $B = 3\,\mu$eV, $A = 0.5\,\mu$eV and $\Gamma = 150\,\mu$eV. \label{fig:bz}}
\end{figure}

When the tunnel couplings are equal, $q_1 = q_2$, Eq.~(\ref{eq:Ibz}) yields a Lorentzian dip at resonance, see Fig.~\ref{fig:bz}(a), where we compare the approximate analytic result with our numerical simulations.
The dip in the current around $\delta = 0$ can be understood by realizing that the resonant driving now allows for sequential elastic photon-assisted transitions $\ket 1 \to \ket S \to \ket 2$ (and vice versa), which slightly reduces the probability over time to end up in the decaying state $\ket S$.
When the two tunnel couplings become different, $q_1 \neq q_2$, we see that Eq.~(\ref{eq:Ibz}) no longer describes a Lorentzian, but becomes asymmetric, showing both regions with increased and decreased current as compared to the background, see Fig.~\ref{fig:bz}(b,c).
We interpret this as being the result of competition between the mechanism outlined above (resulting in a decrease of the current) and the enhancement of the current caused by the coherent mixing of the two (1,1) states for unequal couplings, as explained in the previous Section, where the Rabi driving is now mediated via the exchange coupling.

Fig.~\ref{fig:bz} shows that in all cases our simple analytic expressions match all features of the line shape very well.
Comparing these results with those of Sec.~\ref{sec:bx}, we see that our analytic insight can thus indeed reveal qualitative information about the mechanisms underlying the EDSR observed in an experiment, based on analyzing the line shape of the resonant current response.

\subsection{Effective electric driving}

In the previous sections we took the $3\times 3$ master equation of Eq.~(\ref{eq:lind3}) and separated the system into the (1,1) subspace with slow dynamics, containing the driving, and the remaining state $\ket{S}$ with fast dynamics, set by the decay rate $\Gamma$.
However, when the driving couples most strongly to the energy of $\ket{S}$, such separation of time scales does not work because the driving element now lies within the \textit{fast} subspace.
In this case we evaluate the current by first calculating the transition rates $\Gamma_{ij}$ between the states $i,j$ in the set $\{ 1,2,S \}$. We then set up a classical master equation for the steady-state occupation probabilities $p_n$ of the three levels,
\begin{align}
    -\sum_{f} \Gamma_{fn}p_n + \sum_i \Gamma_{ni} p_i = 0,
\end{align}
which, together with the boundary condition $\sum_n p_n = 1$, yields a solution from which we can calculate the current as $I = \Gamma p_s$~\cite{Note1}.

We evaluate the rates $\Gamma_{ij}$ using a Fermi-golden-rule-like approach, similar to that used in Ref.~\cite{Danon2014}, where we expand the resulting oscillating occupation probabilities to lowest order in $A/\Gamma$, see App.~\ref{app:FGR} for the details.
This yields a current that we can write as
\begin{align}\label{eq:Idet}
    \frac{I}{I_0} = {} & {} 1 - \frac{4A^2}{4E_0^2+\Gamma^2} \bigg[\frac{2E_0^2}{4E_0^2+\Gamma} \frac{\gamma_1^2+\gamma_2^2}{\gamma^2 + \delta^2} \nonumber\\
    {} & {} \quad - \frac{4E_0^2+\Gamma^2}{16 \Gamma^2}\frac{(\gamma_1-\gamma_2)^2}{\gamma^2 + \delta^2} \nonumber\\
    {} & {} \quad + \frac{4E_0^2-\Gamma^2}{4E_0^2+\Gamma^2} \frac{(\epsilon_2-\epsilon_1)\delta}{\gamma^2 + \delta^2} \bigg],
\end{align}
again assuming the limit $\Gamma \gg A, B, q_{1,2}$.

\begin{figure}
  \centering
  \includegraphics[width=\linewidth]{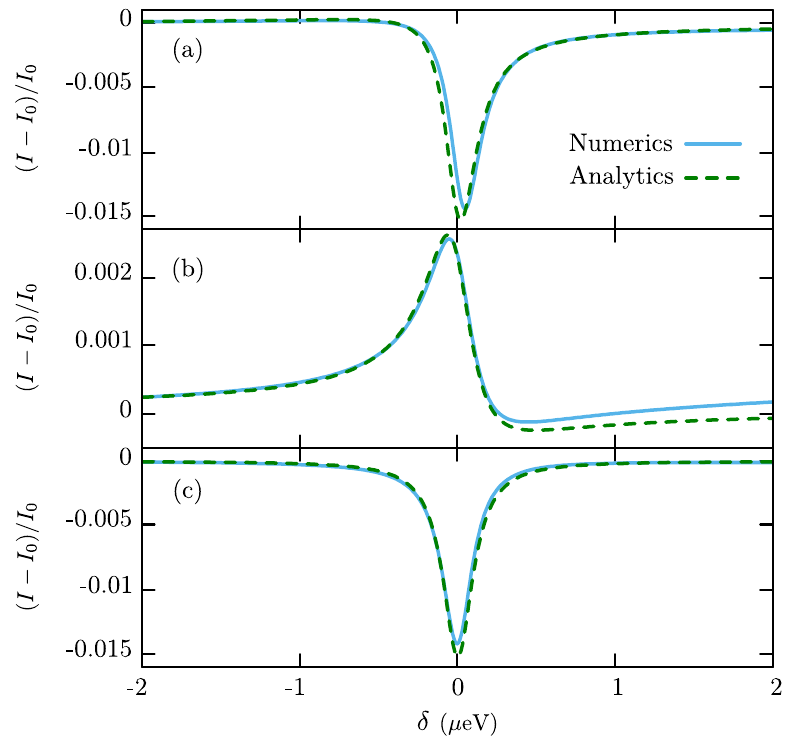}
  \caption{Current profile for a driving in the detuning of the outgoing quantum dot. The dashed green curves are obtained from Eq.~(\ref{eq:Idet}), and the numerical results (solid blue) are added for comparison. (a) A difference in the couplings, $q_1 = 3\,\mu$eV and $q_2 = 2\,\mu$eV, and large static detuning $E_0 = 50\,\mu$eV makes the dip in the current asymmetric. (b) When $E_0$ is small, the dip becomes a peak. In this case we have used $E_0 = 5\,\mu$eV, and the same couplings $q_1$ and $q_2$ as before. (c) Equal couplings, $q_1 = q_2 = 2.5\,\mu$eV and a large $E_0 = 50\,\mu $eV yield a much more symmetric dip. The rest of the parameters are $B = 3\,\mu$eV, $A = 20\,\mu$eV and $\Gamma = 150\,\mu$eV, as in Fig.~\ref{fig:num}. \label{fig:d}}
\end{figure}
In Fig.~\ref{fig:d} we compare this approximate analytic result with our numerical simulations.
We again find that for a large range of parameters our analytic expression captures the line shape of the response very well.
Similar to the case of effective magnetic driving along $\hat z$, we find a symmetric dip in the current when the two tunnel couplings are equal, which develops for unequal couplings into an asymmetric line shape that can both look more like a peak or a dip.
The mechanisms underlying this behavior must also be similar to the case of magnetic driving along $\hat z$:
In both cases the driving enables photon-assisted tunneling between the (1,1) and (0,2) subspaces, which is indeed in contrast with the case of magnetic driving along $\hat x$, where photon-assisted processes are mainly enabled \textit{within} the (1,1) subspace.

\section{Discussion}\label{sec:conclusion}

We have studied a model to characterize EDSR experiments and to understand the line shapes of the resulting current resonances when performed in a double quantum dot tuned to the regime of Pauli spin blockade.
The model assumes a system with large intrinsic or artificial SOI, its main effects in the quantum dot system being: (\textit{i}) the emergence of non-spin-conserving tunnel couplings between the (1,1) and the (0,2) subspaces and (\textit{ii}) electric driving of the system resulting effectively in oscillating local magnetic fields.

Our analysis shows that the underlying nature of the SOI-mediated driving changes qualitatively the shape of the resonant current response.
When we assume the SOI to mainly result in an oscillating effective magnetic field along $\hat x$, then Eq.~(\ref{eq:Ibx}) as well as Fig.~\ref{fig:bx} show that this mostly results in an increase of the current at resonance, but only a significant one when the two states at resonance have different coupling strengths to the outgoing (0,2) state.
Indeed, this type of driving induces Rabi rotations within the (1,1) subspace, and a difference between the two escape rates $\gamma_1$ and $\gamma_2$ of the (1,1) states can cause this mixing to increase the current by opening up the ``bottleneck'' formed by the (1,1) state with the slowest escape rate.
When the driving results in an oscillating effective field along $\hat z$ and the system is brought into resonance, photon-assisted transitions between the (1,1) states and the (0,2) state result in a slight decrease of the current, as seen from in Eq.~(\ref{eq:Ibz}) and shown in Fig.~\ref{fig:bz}(a).
For different couplings, $q_1 \neq q_2$, a mechanism similar to the one discussed above and this one compete, resulting in an asymmetric line shape that can resemble both a peak and a dip, as shown in Fig.~\ref{fig:bz}(b,c).

Considering these qualitative differences between the cases of the resulting driving field being parallel or perpendicular to the static background field, our results could thus be used to obtain information about the direction of the (effective) SOI fields in the system, based on the line shape of the EDSR response.
For example, if we compare the EDSR line shapes reported in Refs.~\cite{Crippa2018,Watzinger2018}, we see that the resonance in Ref.~\cite{Crippa2018} (Fig.~2a) looks more symmetric than the one in Ref.~\cite{Watzinger2018} (Fig.~3c), which could indicate that Ref.~\cite{Crippa2018} has an effective SOI field closer to being perpendicular to the applied field than Ref.~\cite{Watzinger2018} (which, at first sight, could be consistent with the higher quality of the Rabi oscillations observed in Ref.~\cite{Crippa2018}).


A driving that mainly couples to the detuning between the (1,1) and (0,2) subspaces produces line shapes similar to the case of effective magnetic driving along $\hat z$, based on a similar competition between a reduced population of $\ket S$ and a mixing-induced increased escape from the (1,1) state when $q_1 \neq q_2$.
Indeed, the currents described by Eqs.~(\ref{eq:Ibz}) and (\ref{eq:Idet}) have a similar form.
There is, nevertheless, a difference: When we consider an oscillating field along $\hat z$, the mechanism that produces an increase of the current is weak, whereas when we consider oscillations in the detuning, this results in an oscillating exchange energy that effectively couples the two (1,1) states in a similar way as when we consider an oscillating magnetic field along $\hat x$.
This effect is much stronger than the mechanism that blocks the current and thus dominates, producing the much more prominent peak of Fig.~\ref{fig:d}(b).

Our analytical results, validated by the good agreement with the numerics, give insight in the connection between EDSR-induced current resonances and the underlying physical mechanisms.
The simple, yet detailed relation between the line shapes and the intrinsic or effective system parameters that we present in this paper thus provides a means to extract valuable information about the system.
In a standard EDSR experiment, from which usually only spectroscopic information is obtained, our results could be used to gain additional knowledge about the intricate details of the underlying SOI in the system and its effective manifestation.

\textit{Acknowledgements.}---This work is partly supported by the Centers of Excellence funding scheme of the RCN, project number 262633, QuSpin.

\appendix

\begin{widetext}

\section{Time-evolution matrices\label{app:magdrive}}

The matrix ${\cal M}$ used in Eqs.~(\ref{eq:rho2}) and (\ref{eq:rho2F}) reads explicitly as
\begin{align}
	\mathcal{M} = \left( 
	\begin{array}{c c c c}
		-\theta_{11}\Gamma & 2i \theta_{12}E_0 & -2i\theta_{12}E_0 & \theta_{22}\Gamma \\
		-\theta_{12}(2E_0+i\Gamma) & -2i\mathcal B & 0 & \theta_{12}(2E_0-i\Gamma) \\
		\theta_{12}(2E_0-i\Gamma) & 0 & 2i\mathcal B^* & -\theta_{12}(2E_0+i\Gamma) \\
		\theta_{11}\Gamma & -2i\theta_{12}E_0 & 2i\theta_{12}E_0 & -\theta_{22}\Gamma
	\end{array}
	\right),
\end{align}
with $\mathcal B = B+\theta_{11}(E_0-\frac{i}{2}\Gamma) - \theta_{22}(E_0+\frac{i}{2}\Gamma)$, where we introduced the notation
\begin{align}
	\theta_{\alpha\beta} = \frac{2q_\alpha q_\beta}{4E_0^2+\Gamma^2},
\end{align}
and the two driving matrices read as
\begin{align}
	\mathcal{V}_\text{drive}^\text{z} = \left( 
	\begin{array}{c c c c}
		0  &  0  &  0  &  0  \\
		0  &  2i  &  0  &  0  \\
		0  &  0  & -2i  &  0  \\
		0  &  0  &  0  &  0  
	\end{array}
	\right)
	\qquad\text{and}\qquad
	\mathcal{V}_\text{drive}^\text{x} = \left( 
	\begin{array}{c c c c}
		0  &  i  & -i  &  0  \\
		i  &  0  &  0  & -i  \\
		-i  &  0  &  0  &  i  \\
		0  & -i  &  i  &  0  
	\end{array}
	\right).
\end{align}
Without loss of generality for the three-level setup, we assumed both $q_1$ and $q_2$ to be real from here on.
\end{widetext}

\section{Calculation of transition rates\label{app:FGR}}

We now present our calculation of the transition rates $\Gamma_{if}$ between the three levels under effective driving along the detuning, assuming that the singlet decay rate $\Gamma$ is the largest energy scale involved.
We derive a Fermi's golden rule for the time-dependent unperturbed Hamiltonian, which we then expand close to resonance $\omega \approx 2B$ to leading order in $A/\Gamma$.

We split the Hamiltonian in two parts, $H = H_0 (t) + H_1$,
\begin{align}
H_0(t) = {} & {} B \ket{1}\bra{1} - B\ket{2}\bra{2} \nonumber\\
{} & {} -[ E_0 - A\cos(\omega t)]\ket{S}\bra{S},\label{eq:apph0}
\end{align}
and
\begin{align}
H_1 = q_1 \ket{1}\bra{S} + q_2 \ket{2}\bra{S} + {\rm H.c.},
\end{align}
where we will treat $H_1$ as a perturbation.

We then transform to a time-dependent interaction picture where the perturbation Hamiltonian becomes
\begin{align}
\tilde H_1(t) = {} & {} q_1 e^{-i\int_0^{t} dt'\, [E_S(t')-E_1(t')]} \ket{1}\bra{S}\nonumber\\
{} & {}  + q_2 e^{-i\int_0^{t} dt'\, [E_S(t')-E_2(t')]}  \ket{2}\bra{S} + {\rm H.c.},
\end{align}
with $E_n(t)$ the time-dependent eigenenergies set by (\ref{eq:apph0}).
In this picture the matrix elements of the time-evolution operator can formally be expressed as
\begin{align}
\bra f {} & {} U(t,0)\ket i  =  \braket{f|i} \nonumber\\
            {}&{} -i\int_{0}^t dt_1 \bmk{f}{H_1}{i}e^{-i\int_0^{t_1} dt_2\, [E_i(t_2)-E_f(t_2)]} \notag \\
            {}&{}       -\int_{0}^t dt_2 \int_{0}^{t_2} dt_1 \sum_{k}  \bmk{f}{H_1}{k}\bmk{k}{H_1}{i} \notag \\
            {}&{}       \quad\times e^{-i\int_0^{t_1} dt_3 \int_0^{t_2} dt_4\, [E_i(t_3)-E_k(t_3)+E_k(t_4)-E_f(t_4)]} \notag \\
            {}&{}  + \mathcal{O}(q_{1,2}^3),
\end{align}
from which we can calculate transition rates as
\begin{align}
\Gamma_{if} = \frac{d}{dt}| \bra f U(t,0)\ket i |^2.\label{eq:gammaif}
\end{align}

\begin{figure}
	\centering
	\includegraphics[width=\linewidth]{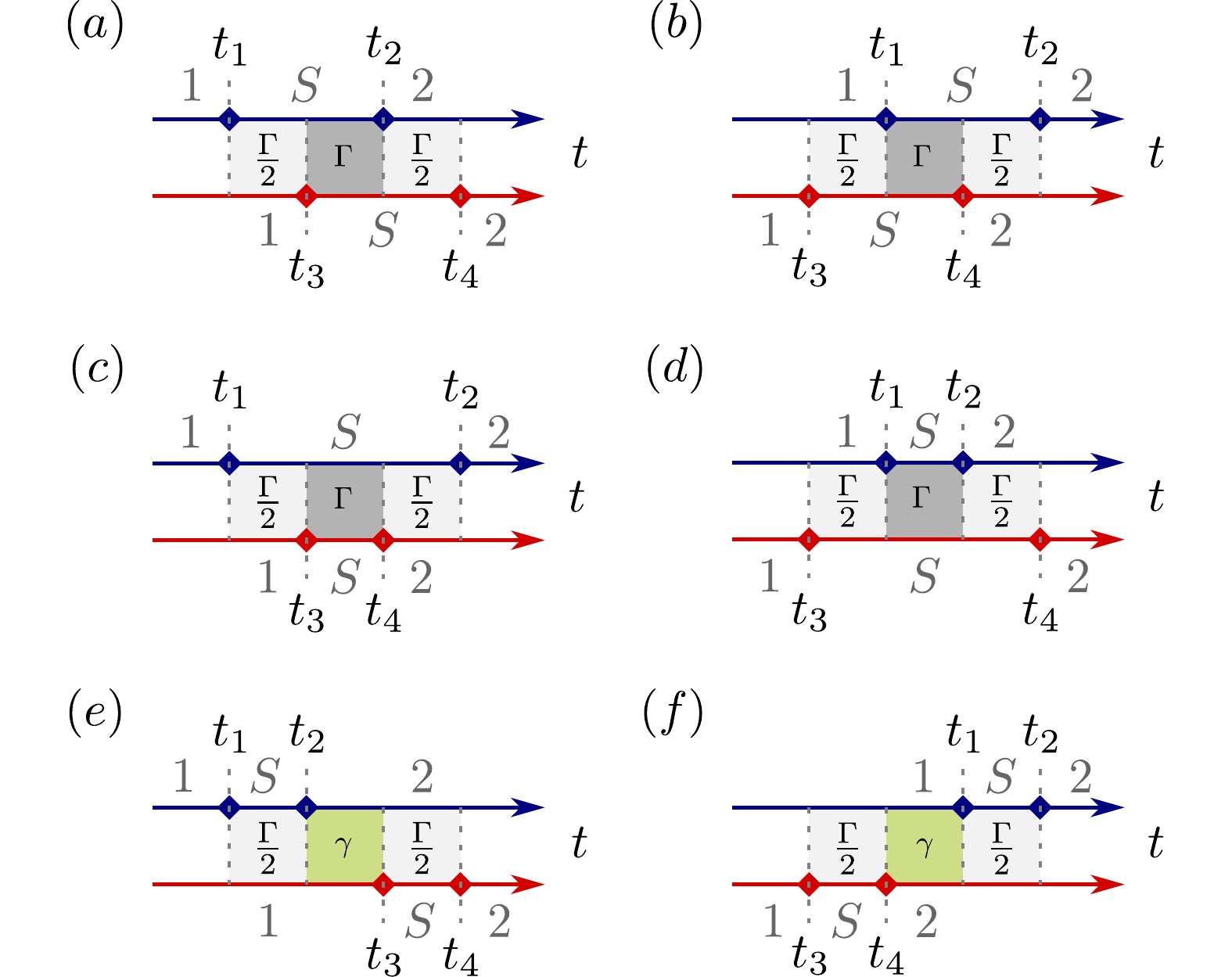}
	\caption{All possible time orderings in Eq.~(\ref{eq:c2_2}) where the blue (red) arrow represents forward (backward) propagation. The different shadings indicate the different effective decay rates we included in the evaluation.\label{fig:diagram1}}
\end{figure}

First we will focus on the transition rate $\Gamma_{12}$, that is, the rate from the state $\ket 1$ to $\ket 2$.
This will also allow us to show how we include the effect of the escape rate $\Gamma$ into our expressions.
The lowest-order processes contributing to this rate has both its forward- and backward-propagating time-evolution operator in (\ref{eq:gammaif}) of second order in $q_{1,2}$.
Explicitly, the approach outlined above yields for this contribution to the transition probability $| \bra 2 U(t,0)\ket 1 |^2$ the expression
\begin{align}
{} & {} q_1^2 q_2^2 \int_{0}^t dt_4 \int_{0}^{t_4} dt_3 \int_{0}^{t} dt_2 \int_{0}^{t_2} dt_1  \nonumber\\
{} & {} \quad\times e^{-i(B+E_0)(t_1-t_3)} e^{-i(B-E_0)(t_2-t_4)}   \nonumber\\
{} & {} \quad\times  e^{-i\frac{A}{\omega} [-\sin(\omega t_1) + \sin(\omega t_2) + \sin(\omega t_3) - \sin(\omega t_4)]}.\label{eq:c2_2}
\end{align}
The six possible relative orderings of the times in this expression are illustrated by the diagrams shown in Fig.~\ref{fig:diagram1}, where the blue (red) arrows depict forward-propagating (backward-propagating) time evolution.

We first focus on diagrams (a--d), which describe the sequential process $\ket 1 \to \ket S \to \ket 2$.
Indeed, all four diagrams contain a ``full'' transition from $\ket 1$ to $\ket S$, after which the system remains in $\ket S$ for some time [e.g., $t_2-t_3$ in diagram (a)], followed by a transition to $\ket 2$.

We then account for the decay of the singlet state $\ket S$ by including a factor $e^{-\frac{1}{2}\Gamma (t_2-t_1)}e^{-\frac{1}{2}\Gamma (t_4-t_3)}$ in the integrand, which is the same factor that would result from adding an imaginary part to the eigenenergy $E_S(t) \to E_S(t) -\frac{i}{2}\Gamma$.
In other words, one could say that we switched to a perturbative expansion of a Lindblad equation instead of a Schr\"{o}dinger equation, where we included decay of the element $\ket S\bra S$ with rate $\Gamma$ (indicated with dark gray in the diagrams) and decay of the interference terms $\ket S\bra n$ and $\ket n\bra S$ with rate $\Gamma/2$ (light gray).

Using that diagrams (b,d,f) are the complex conjugate of (a,c,e) this finally yields for (a,b) the expression
\begin{align}
{\rm (a)} {} & {} + {\rm (b)} =\nonumber\\
	{} & {} 2 q_1^2 q_2^2 {\rm Re} \bigg\{ \int_{0}^t dt_4 \int_{0}^{t_4} dt_2 \int_{0}^{t_2} dt_3 \int_{0}^{t_3} dt_1  \nonumber\\
	{} & {} \quad\times e^{-i(B+E_0)(t_1-t_3)} e^{-i(B-E_0)(t_2-t_4)}   \nonumber\\
	{} & {} \quad\times  e^{-i\frac{A}{\omega}  [-\sin(\omega t_1) + \sin(\omega t_2) + \sin(\omega t_3) - \sin(\omega t_4)]} \nonumber\\
	{} & {} \quad\times e^{-\frac{1}{2}\Gamma (t_3-t_1)}e^{-\Gamma(t_2-t_3)}e^{-\frac{1}{2}\Gamma (t_4-t_2)} \Big\}.\label{eq:c2_2ab}
\end{align}
We introduce the time-difference coordinates $\tau_{ij} = t_i - t_j$ and, since $\Gamma$ is the largest energy scale involved, we anticipate that the integrand vanishes rapidly for increasing $\tau_{31}$, $\tau_{23}$, or $\tau_{42}$.
We then complete our set of coordinates with the average time $\sigma= \frac{1}{4}(t_1+t_2+t_3+t_4)$, transform the integral to the new coordinate system, linearize the oscillating exponentials in the small time differences,
\begin{align}
	\frac{A}{\omega} [- {} & {} \sin(\omega t_1) + \sin(\omega t_2) + \sin(\omega t_3) - \sin(\omega t_4)] \approx \nonumber\\
	{} & {} A(\tau_{31} - \tau_{42})\cos(\omega\sigma),
\end{align}
and extend the upper limit of integration for the difference coordinates to infinity, for convenience.
This finally yields the approximate expression
\begin{align}
	{\rm (a)} {} & {} + {\rm (b)} =\nonumber\\
	{} & {} 2 q_1^2 q_2^2 {\rm Re} \bigg\{ \int_{0}^t d\sigma \int_{0}^{\infty} d\tau_{42} \int_{0}^{\infty} d\tau_{23} \int_{0}^{\infty} d\tau_{31}  \nonumber\\
	{} & {} \quad\times e^{[iB+iE_0-iA\cos(\omega\sigma)-\frac{1}{2}\Gamma]\tau_{31}}  \nonumber\\
	{} & {} \quad\times  e^{[iB-iE_0+iA\cos(\omega\sigma)-\frac{1}{2}\Gamma]\tau_{42}} e^{-\Gamma\tau_{23}} \Big\},\label{eq:c2_2ab2}
\end{align}
which can be evaluated easily.
The contribution of diagrams (c,d) follows in a similar manner, and we find for the total contribution of the first four diagrams
\begin{align}
	\Gamma_{12}^{\text{(a--d)}} =  \frac{16 q_1^2 q_2^2 \Gamma}{16 \Delta_-^4 + 8\Delta_+^2\Gamma^2 + \Gamma^4},
\end{align}
using the notation $\Delta_\pm^2 = B^2\pm[E_0-A\cos(\omega t)]^2$.
Assuming $B \ll E_0$ for simplicity (which is in accordance with all regimes we investigate in the main text), this yields to lowest order in $A/\Gamma$
\begin{align}
	\Gamma_{12}^{\text{(a--d)}} \approx  \frac{16 q_1^2 q_2^2 \Gamma}{(4E_0^2 + \Gamma^2)^2}.\label{eq:contad}
\end{align}
This result can be understood since these four diagrams describe sequential tunneling:
Transitions from $\ket 1$ to $\ket S$ happen with a rate $\gamma_1$, after which a transition from $\ket S$ to $\ket 2$ rapidly follows, before $\ket S$ decays, with a probability $\gamma_2/\Gamma$.
The total rate is thus expected to be approximately $\gamma_1\gamma_2/\Gamma$, which is identical to the result (\ref{eq:contad}).

We now proceed to investigate diagrams (e,f), which contain a period of time $\tau$ (marked green in Fig.~\ref{fig:diagram1}) during which the states $\ket{1}$ and $\ket{2}$ interfere, resulting in a coherence factor of the form $e^{\pm i 2B\tau}$ due to their energy difference of $2B$.
In the presence of harmonic driving, these coherence factors can thus result in a resonant response, when $\omega \approx 2B$.
In order to capture the correct line shape of this response, we need to include decoherence between the levels $\ket 1$ and $\ket 2$, in our model dominated by escape to the drain lead with rates $\gamma_1$ and $\gamma_2$, respectively.

Using again that diagrams (e) and (f) are each other's complex conjugate, we thus write
\begin{align}
	{\rm (e)} {} & {} + {\rm (f)} =\nonumber\\
	{} & {} 2 q_1^2 q_2^2 {\rm Re} \bigg\{ \int_{0}^t dt_4 \int_{0}^{t_4} dt_3 \int_{0}^{t_3} dt_2 \int_{0}^{t_2} dt_1  \nonumber\\
	{} & {} \quad\times e^{-i(B+E_0)(t_1-t_3)} e^{-i(B-E_0)(t_2-t_4)}   \nonumber\\
	{} & {} \quad\times  e^{-i\frac{A}{\omega}  [-\sin(\omega t_1) + \sin(\omega t_2) + \sin(\omega t_3) - \sin(\omega t_4)]} \nonumber\\
	{} & {} \quad\times e^{-\frac{1}{2}\Gamma (t_2-t_1)}e^{-\gamma(t_3-t_2)}e^{-\frac{1}{2}\Gamma (t_4-t_3)} \Big\}.\label{eq:c2_2ef}
\end{align}
Now there are only two time differences that will always be small, $\tau_{21}$ and $\tau_{43}$, and we complement these two coordinates with the averages $\sigma_{21} = \frac{1}{2}(t_2+t_1)$ and $\sigma_{43} = \frac{1}{2}(t_4+t_3)$.
Then, following the same steps as before, we arrive at
\begin{align}
	{\rm (e)} {} & {} + {\rm (f)} = \nonumber\\
	{} & {} 2 q_1^2 q_2^2 \text{Re} \bigg\{ \int_{0}^t d\sigma_{43} \int_{0}^{\sigma_{43}} d\sigma_{21} \int_{0}^{\infty} d\tau_{43} \int_{0}^{\infty} d\tau_{21} \nonumber\\
    {} & {} \quad   \times e^{i2B(\sigma_{43}-\sigma_{21})} e^{iE_0(\tau_{21}-\tau_{43})} \nonumber  \\
    {} & {} \quad   \times  e^{- i A [\cos(\omega \sigma_{21})\tau_{21} - \cos(\omega \sigma_{43})\tau_{43}]}  \nonumber \\
    {} & {} \quad   \times e^{-\frac{\Gamma}{2} (\tau_{21}+\tau_{43})} e^{-\gamma (\sigma_{43}-\sigma_{21})}\Big\},\label{eq:c2_efa}
\end{align}
where we used $\sigma_{43}-\frac{1}{2}\tau_{43} \approx \sigma_{43}$.
Integrating over the two time differences and taking the time derivative yields a contribution to the transition rate of
\begin{align}
	{} & {} \Gamma_{12}^{{\rm (e,f)}}  = 2 q_1^2 q_2^2 \text{Re} \bigg\{ \int_{0}^{t} d\sigma_{21} \, e^{(i2B-\gamma)(t-\sigma_{21})} \nonumber\\
	{} & {} \quad\times  \frac{1}{E_0+i\tfrac{\Gamma}{2} + A \cos( \omega \sigma_{21})} \frac{1}{E_0-i\tfrac{\Gamma}{2} + A \cos( \omega t)}\bigg\},\label{eq:c2_efb}
\end{align}
We then expand both fractions in small $A/\Gamma$, yielding
\begin{align}
 \frac{1}{E_0 \pm i\tfrac{\Gamma}{2} + A \cos( \omega t)} \approx
 \frac{1}{E_0 \pm i\tfrac{\Gamma}{2}} - \frac{A(e^{i\omega t} + e^{-i\omega t})}{2(E_0 \pm i\tfrac{\Gamma}{2})^2}.
\end{align}
Focusing on the single-photon resonance where $\omega \approx 2B$ we discard all time-dependent terms except the one proportional to $A^2 e^{i\omega(\sigma_{21}-t)}$ (effectively thus using a rotating wave approximation), which results in
\begin{align}
	\Gamma_{12}^{{\rm (e,f)}}  = {} & {} \frac{8 q_1^2 q_2^2}{4E_0^2 + \Gamma^2}\text{Re} \bigg\{ \int_{0}^{t} d\tau \, e^{i2B\tau-\gamma\tau} \bigg\}  \nonumber\\
	{} & {}  + \frac{8A^2q_1^2 q_2^2}{(4E_0^2 + \Gamma^2)^2}\text{Re} \bigg\{ \int_{0}^{t} d\tau \, e^{i(2B-\omega)\tau-\gamma\tau} \bigg\},\label{eq:c2_efc}
\end{align}
which converges for $t \gtrsim \gamma^{-1}$ to
\begin{align}
 \Gamma_{12}^{{\rm (e,f)}} \approx \frac{ 8 q_1^2 q_2^2}{4E_0^2 + \Gamma^2}  \frac{\gamma}{4 B^2 + \gamma^2} + \frac{8A^2 q_1^2 q_2^2}{(4E_0^2 + \Gamma^2)^2 }  \frac{\gamma}{\delta^2 + \gamma^2},\label{eq:contef}
\end{align}
where we used the notation $\delta = \omega-2B$ again.

Combining Eqs.~(\ref{eq:contad}) and (\ref{eq:contef}) we find the total rate,
\begin{align}
	\Gamma_{12} \approx  \frac{16 q_1^2 q_2^2 \Gamma}{(4E_0^2 + \Gamma^2)^2}
	\left( 1 + \frac{q_1^2+q_2^2}{4B^2 + \gamma^2} + \frac{A^2}{2\Gamma} \frac{\gamma}{\delta^2 + \gamma^2}\right).
\end{align}
The opposite rate $\Gamma_{21}$ follows from swapping $q_1 \leftrightarrow q_2$ and changing the signs of $B$ and $\omega$, which leaves the expression unchanged, indicating that the rates are equal, within the approximations used.

\begin{figure}
	\centering
	\includegraphics[width=\linewidth]{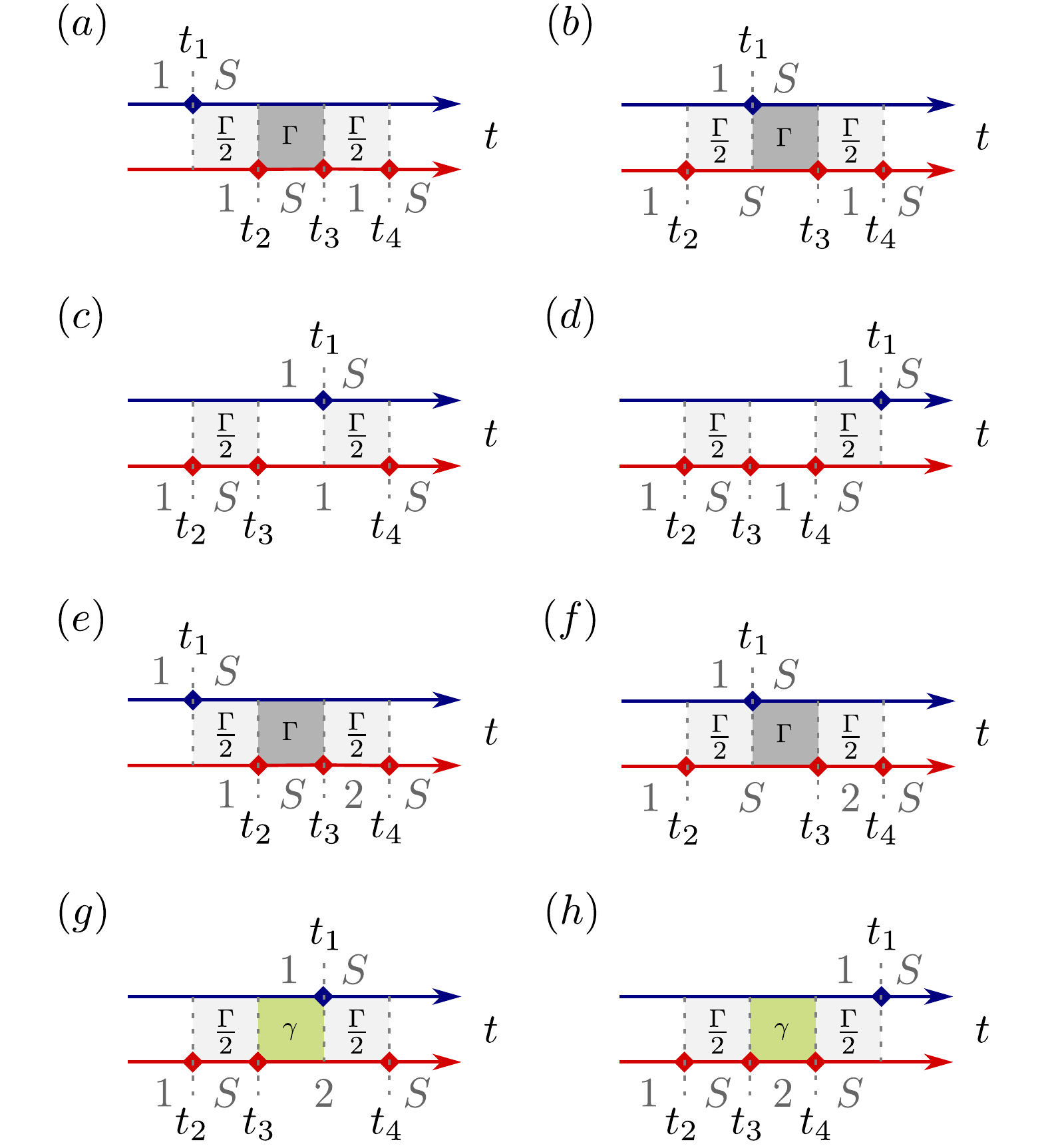}
	\caption{All possible time orderings for the two processes included in Eq.~(\ref{eq:cs_2}). Only diagrams (g,h) can contribute to a resonant response due to finite interference between time spent in the states $\ket 1$ and $\ket 2$.\label{fig:diagram2}}
\end{figure}

Now we consider the transition rates $\Gamma_{1S}$ and $\Gamma_{2S}$.
The dominating contributions are simply $\gamma_{1,2}$, which follow from second-order perturbation theory in $q_{1,2}$.
Focusing on the rate $\Gamma_{1S}$, we see that the next correction is of fourth order and yields contributions to the transition probability stemming from two different processes,
\begin{align}
	-{} & {} 2 q_1^2 q_2^2 {\rm Re} \bigg\{ \int_{0}^t dt_4 \int_{0}^{t_4} dt_3 \int_{0}^{t_3} dt_2 \int_{0}^{t} dt_1  \nonumber\\
	{} & {} \quad\times \Big( e^{-i(B+E_0)(t_1-t_2)} e^{-i(-B-E_0)(t_4-t_3)}   \nonumber\\
	{} & {} \qquad\times  e^{-i\frac{A}{\omega} [-\sin(\omega t_1) + \sin(\omega t_2) - \sin(\omega t_3) + \sin(\omega t_4)]} \nonumber\\
	{} & {} \quad+ e^{-i(B+E_0)(t_1-t_2)} e^{-i(B-E_0)(t_4-t_3)}   \nonumber\\
	{} & {} \qquad\times  e^{-i\frac{A}{\omega} [-\sin(\omega t_1) + \sin(\omega t_2) - \sin(\omega t_3) + \sin(\omega t_4)]}\Big)\Big\},\label{eq:cs_2}
\end{align}
where the minus sign in front is due to the fact that both the first-order contribution to $U(t,0)$ and the third-order contribution to $U(t,0)^\dagger$ come with a factor $-i$.
All possible time orderings in (\ref{eq:cs_2}) are indicated in Fig.~\ref{fig:diagram2}; the first term within the brackets in (\ref{eq:cs_2}) corresponds to diagrams (a--d) and the second one to (e--h).
We then see that all six first diagrams (a--f) present a small correction to the (second-order) rate $\gamma_1$ and will not yield a resonant response; we therefore neglect their small contribution and focus solely on diagrams (g,h),
\begin{align}
	{\rm (g)} {} & {} + {\rm (h)} = \nonumber\\
	-{} & {} 4 q_1^2 q_2^2 {\rm Re} \bigg\{ \int_{0}^t d\sigma_{41} \int_{0}^{\sigma_{41}} d\sigma_{32} \int_{0}^{\infty} d\tau_{32}  \nonumber\\
	{} & {} \hspace{5em}\times e^{(-i2B-\gamma)(\sigma_{41}-\sigma_{32})}  \nonumber\\
	{} & {} \hspace{5em}\times  e^{[-iE_0+iA\cos(\omega\sigma_{32})-\frac{1}{2}\Gamma]\tau_{32}}\Big\} \nonumber\\
	{} & {} \quad\times   {\rm Re} \bigg\{ \int_{0}^{\infty} d\tau_{41} e^{[iE_0-iA\cos(\omega\sigma_{41})-\frac{1}{2}\Gamma] \tau_{41}}\bigg\},\label{eq:cs_2a}
\end{align}
where we made the same approximations as before.
This finally yields a contribution to the transition rate of
\begin{align}
	\Gamma_{1S}^{{\rm(g,h)}} = {} & {}  \frac{8 q_1^2 q_2^2\Gamma}{4[E_0-A\cos(\omega t)]^2 + \Gamma^2}\nonumber\\
	{} & {}  \times{\rm Re} \bigg\{ \int_{0}^{t} d\sigma_{32}\frac{e^{(-i2B-\gamma)(t-\sigma_{32})}}{-iE_0+iA\cos(\omega\sigma_{32})-\frac{1}{2}\Gamma}\bigg\}.\label{eq:cs_2b}
\end{align}

We then again expand to lowest order in $A/\Gamma$, keeping only the time-dependent terms that contribute to the resonant response and discard all the non-resonant contributions, which present a small correction to the second-order rate $\gamma_1$.
This yields for $t \gtrsim \gamma^{-1}$ the total rate

\begin{align}\label{eq:gS_1}
	\Gamma_{1S} \approx {}&{} \gamma_1 -\frac{64 q_1^2q_2^2 A^2 E_0\Gamma [4E_0\Gamma\gamma + (4E_0^2 - \Gamma^2)\delta] }{(4E_0^2+\Gamma^2)^4(\delta^2+\gamma^2)}.
\end{align}

\begin{figure}[b!]
	\centering
	\includegraphics[width=\linewidth]{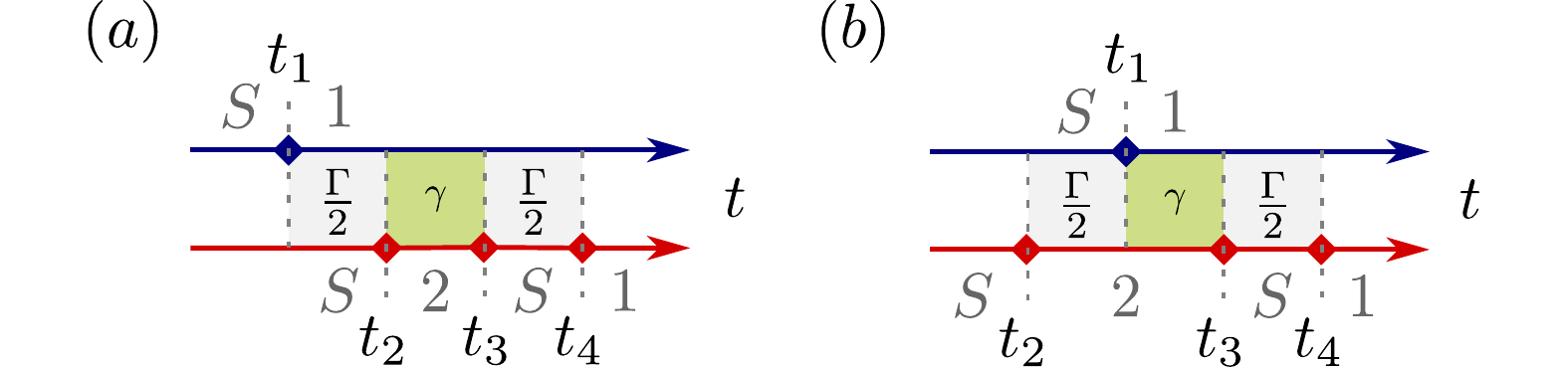}
	\caption{Two fourth-order diagrams that in principle can yield a resonant response in the transition rate $\Gamma_{S1}$.\label{fig:dia_S1}}
\end{figure}

The correction to the rate $\Gamma_{2S}$ follows again by swapping $q_1 \leftrightarrow q_2$ and changing the signs of $B$ and $\omega$, now yielding a slightly different expression,
\begin{align}\label{eq:gS_2}
	\Gamma_{2S} \approx {}&{} \gamma_2-\frac{64 q_1^2q_2^2 A^2 E_0\Gamma [4E_0\Gamma\gamma - (4E_0^2 - \Gamma^2)\delta] }{(4E_0^2+\Gamma^2)^4(\delta^2+\gamma^2)}.
\end{align}

Finally, we can consider the transition rates $\Gamma_{S1}$ and $\Gamma_{S2}$.
The structure of the corresponding diagrams is similar to those shown in Fig.~\ref{fig:diagram2}; in Fig.~\ref{fig:dia_S1} we show the two contributions to $\Gamma_{S1}$ that in principle can yield a resonant response.
Closer inspection of the structure of these diagrams reveals that these contributions add up to a correction to the rate that is equal to $\Gamma_{1S}^{{\rm (g,h)}}$ evaluated above, thus yielding
\begin{align}
	\Gamma_{S1} \approx {}&{}  -\frac{64 q_1^2q_2^2 A^2 E_0\Gamma [4E_0\Gamma\gamma + (4E_0^2 - \Gamma^2)\delta] }{(4E_0^2+\Gamma^2)^4(\delta^2+\gamma^2)}, \\
	\Gamma_{S2} \approx {}&{} -\frac{64 q_1^2q_2^2 A^2 E_0\Gamma [4E_0\Gamma\gamma - (4E_0^2 - \Gamma^2)\delta] }{(4E_0^2+\Gamma^2)^4(\delta^2+\gamma^2)}.
\end{align}
Since $\ket S$ also decays to the drain lead with rate $\Gamma$, the importance of $\Gamma_{S1}$ and $\Gamma_{S2}$ for the total dynamics of the three-level system is a factor $\sim q_{1,2}^2/\Gamma^2$ smaller than that of the resonant corrections in the rates $\Gamma_{1S}$ and $\Gamma_{2S}$.
For this reason we can neglect $\Gamma_{S1}$ and $\Gamma_{S2}$.

%

\end{document}